\documentclass[12pt]{article}
\begin{document}

\title{ON THE PROBLEM OF COSMOLOGICAL SINGULARITY WITHIN GAUGE THEORIES OF GRAVITATION}

\author{G.V. VERESHCHAGIN \\
{\it \small I.C.R.A. - International Center for Relativistic Astrophysics,} \\
{\it \small University of Rome "La Sapienza", Physics Department,} \\
{\it \small P.le A. Moro 5, 00185 Rome, Italy.} \\
{\it \small Email: veresh@icra.it}}

\date{}

\maketitle

\begin{abstract}
In this note I discuss the problem of cosmological singularities within gauge theories of gravitation. Solutions of cosmological equations with the scalar field are considered.
\end{abstract}

The problem of cosmological singularity is one of the most fundamental issues in modern theoretical cosmology. Investigation of this problem within gauge theories of gravitation (GTG) is of undoubted interest, since numerous attempts of resolution of this problem did not reach success within general relativity (GR). Generalized cosmological Freidmann equations (GCFE) were derived in 1980 in \cite{Min80} within the framework of GTG. It possesses important regularizing properties and allows obtaining nonsingular solutions having friedmannian asymptotics of GR.

The study of cosmological models with the scalar field within GTG is motivated by successes of existing inflationary models within GR. The first regular solution with nonlinear scalar field was obtained in \cite{Min01}. Since analytical solution of GCFE does not exist, the qualitative analysis (by analogy with GR \cite{Bel85}) seems to be promising. It turns out that unlike GR the phase space of the dynamical system, corresponding to GCFE in the case of scalar field, is bounded within GTG \cite{Ver}. The boundary of the phase space contains special solutions of cosmological equations \cite{Min}. The presence of this boundary allows avoiding divergencies of the scalar field derivative with respect to time on contraction stage that take place for most solutions within GR in the finite region for the scalar field.

Within GR in simplest cases the stage of cosmological expansion is described by the phase portrait with the stable focus singular point at the origin \cite{Bel85}. On the phase portrait describing contraction stage the corresponding singular point is unstable focus. Moreover, the two phase portraits are independent. In order to locate the solution one has to determine to sign of the Hubble parameter.

Within GTG the phase space has much more complicated structure. Close to the origin the phase portraits practically coincide with the corresponding portraits within GR since in this case the GCFE is equivalent to the Friedmann equation. However, according to \cite{Min} the Hubble parameter vanishes in the finite region of the phase space within GTG. As a consequence, there are inevitably regular transitions from contraction to expansion and back in the finite region of the phase space even in the case of flat and open type models in addition to close type models within GR. Moreover, the phase space has nontrivial topology: the phase portraits with stable and unstable focuses are glued along the boundaries of the phase space (in the case of negative indefinite parameter of cosmological equations of GTG). This leads to the remarkable consequence that the phase trajectories beginning in unstable focus reach the boundary of the phase space and then may pass to the phase portrait with stable focus and end there, rather than go to infinity. The Hubble parameter is negative inside the unstable focus and positive inside the stable focus (this circumstance can be considered as the consequence of the correspondence with GR). Thus, the phase trajectories cross the curves where the Hubble parameter vanishes and the corresponding solutions describe the regular transition from contraction to expansion. Notice in addition, that most regular solutions incorporate inflationary stage since in the region of the phase space close to the origin (again as a correspondence with GR) there are inflationary separatrices, playing the role of attractors \cite{Bel85}.

In refs. \cite{Ver} the `glued' solutions are considered, being composed of a particular solution and the special (envelope) solution. Existence of such solutions is beyond question from the formal mathematical point of view. Since special solutions of GCFE diverge with simplest effective potentials of the scalar field, the conclusion that the singularity is inevitable for cosmological equations of GTG was obtained \cite{Ver}. However, it was out of account, that transition from the particular solution to the special solution and back is possible only with discontinuous behavior of the second derivative of the scale factor (the first derivative of the Hubble parameter \cite{Min}) with respect to time. From the physical point of view, the assumption about such discontinuous point in the cosmological solution seems to be too strong. As a consequence, the conclusion that most solutions of cosmological equations of GTG are singular \cite{Ver}, based implicitly on this assumption, appears to be a premature conclusion.

Indeed, the requirement that the derivative of the Hubble parameter be a continuous function of time excludes all the singular solutions from the consideration \cite{Min}. In this context, conclusions that `...all cosmological solutions for flat, open and closed models are regular in metrics, Hubble parameter, its time derivative...' and `...GTG permit to build nonsingular cosmology...' \cite{Min} are certainly correct.

The series of such regular solutions of cosmological equations of GTG with the scalar field is considered in my thesis \cite{PhD}. Solutions with discontinuous behavior of the Hubble parameter derivative \cite{Ver} are not considered there, and the question on the physical meaning of such solutions remains open.

\end{document}